\documentclass[12pt]{article}
\usepackage{amsmath,amssymb,amsfonts,amstext}
\usepackage{tabularx}
\usepackage{hyperref}
\usepackage{graphicx}
\usepackage{amsfonts}
\usepackage[usenames]{color}
\definecolor{Blue}{rgb}{0,0.08,0.45}
\definecolor{Magenta}{cmyk}{0.1,0.8,0,0.1}
\definecolor{Orange}{rgb}{1,0.5,0}

\begin{document}
\title{ On gravity localization in scalar braneworlds with a 
super-exponential warp factor} 
\author{ Mariana Carrillo--Gonz\'alez$^{a}$, Gabriel Germ\'an$^{b,}\!\!$
\footnote{Corresponding author: gabriel@fis.unam.mx}\,\,,
Alfredo Herrera--Aguilar$^{b,c,d}$ \\
and\, Dagoberto Malag\'on--Morej\'on$^{b}$\\
\\
{\normalsize \textit{$^a$Facultad de Ciencias,}}
{\normalsize \textit{Universidad Aut\'onoma del Estado de Morelos,}}\\
{\normalsize \textit{Av. Universidad 1001, Col. Chamilpa, CP 62209, Cuernavaca, Morelos, M\'{e}xico.}}\\
{\normalsize \textit{$^b$Instituto de Ciencias F\'isicas,} }
{\normalsize \textit{Universidad Nacional Aut\'onoma de M\'exico,}}\\
{\normalsize \textit{Apdo. Postal 48-3, 62251 Cuernavaca, Morelos, M\'{e}xico.}}\\
{\normalsize \textit{$^c$Mesoamerican Centre for Theoretical Physics,} }
{\normalsize \textit{Universidad Aut\'onoma de Chiapas,}}\\
{\normalsize \textit{Km. 4 Carretera Emiliano Zapata, CP 29000, Tuxtla Guti\'errez, Chiapas, M\'{e}xico.}}\\
{\normalsize \textit{$^d$Instituto de F\'{\i}sica y Matem\'{a}ticas,}}
{\normalsize \textit{Universidad Michoacana de San Nicol\'as de Hidalgo,}}\\
{\normalsize \textit{Edificio C--3, Ciudad Universitaria, CP 58040, Morelia, Michoac\'{a}n, M\'{e}xico.}}}
\date{}
\maketitle

\begin{abstract}

We show that within single brane tachyonic braneworld models, {\it super-exponential} warp factors of
the form $e^{-2f} \sim e^{-2c_1e^{c_2 |\sigma|}}$ are problematic when dealing 
with both the finiteness of the effective four-dimensional (4d) Planck mass and the 
localization of 4d gravity, which can be stated by the requirement that $\int e^{-2f(\sigma)}d\sigma
< \infty$, because this condition necessarily implies that $c_1$ and $c_2$ should be positive. 
As a consequence of this fact the
tachyonic field $T$ turns out to be complex in contradiction with the real nature of the starting
action for the tachyonic braneworld. Conversely if one requires to have a real tachyon field, 4d 
gravity will not be localized and the effective gravitational coupling will be infinite. We present 
several typical examples where this problem occurs: 
we have analysed this situation for thin as well as thick
tachyonic braneworlds with 4d Poincar\'e symmetry, for the case when a bulk
cosmological constant is present, and even for a brane with an induced spatially flat
4d cosmological background, and shown that in all cases the tachyon field $T$ comes
out to be inconsistently complex when imposing localization of 4d gravity on the brane. 
On the other hand, when dealing with a further reduction of the hierarchy problem on a two-brane system,
one should carefully 
consider the sign of the constants $c_1$ and $c_2$ to avoid inconsistencies in the tachyonic braneworld model. 
We also present a similar discusion involving a canonical scalar field in the bulk where none of 
these problems arise and hence, the mass hierarchy and 4d gravity localization problems can be 
successfully addressed at once, i.e., with the same warp factor. Finally, the stability analysis of this scalar tensor braneworld model with
a super-exponential warp factor is performed.
\end{abstract}

\section{Introduction}

There has recently been some interest in braneworld models, in which our universe is embedded in a
spacetime with extra dimensions, as a way to solving the mass hierarchy and 4d
gravity localization problems \cite{gogrs}. It has also become a matter of interest to find
smooth braneworld solutions (for interesting reviews on these issues see e.g.,
\cite{thbrs}--\cite{maartens}). Typically such solutions are obtained by introducing one or
several scalar fields in the bulk and the large variety of scalar fields that can be used to
generate these models gives rise to different scenarios \cite{dewolfe}--\cite{koleykar}. 

Several authors have chosen a tachyonic scalar field in the bulk and address issues like the hierarchy
problem, and localization of gravity and matter fields \cite{Bazeiaetal}--\cite{gadrr}. For instance,
in \cite{koleykar} a further reduction of the hierarchy between the fundamental Planck scale and the compactification scale ($kr_c\approx 5$), when compared to the two-brane Randall-Sundrum model where $kr_c\approx 10$, 
is achieved by using a super-exponential warp factor in a two-brane tachyonic braneworld, rendering a model with parameters of the same order. Moreover, all kinds of matter fields as well as gravity can be localized within this super-exponential tachyonic braneworld as long as one is concerned with a two-brane system, since then the higher dimensional volume is always finite. Troubles with gravity localization arise only when one works in the one-brane model, i.e., when the second brane has been removed to infinity along the fifth dimension, since then the volume along the extra dimension can be infinite. On the other hand, when attempting to achieve a further reduction of the hierarchy by implementing a super-exponential warp factor in the two-brane picture, the tachyonic field can also result complex for some choice of the constants $c_1$ and $c_2$ (see Section \ref{tbw}). Thus care should be taken when looking at this problem within tachyonic braneworld models. We have analized this situation in some detail in a simple case and established the resultant nature of the tachyon field $T$ in Table 1. In general, attempts to solve the highly non--linear field equations give rise to imaginary tachyon field configurations within this framework \cite{koleykar,palkar}.

Here we will show that when the braneworld model is generated by gravity and a tachyonic scalar 
field the conditions for localizing 4d gravity on a {\it single} brane as well as the finiteness of the 4d 
gravitational coupling cannot be fulfilled by proposing super-exponential warp factors of the form $e^{-2f} \sim
e^{-2c_1e^{c_2 |\sigma|}}$. The reason is that the relevant integral that defines both of these conditions,
$\int e^{-2f(\sigma)}d\sigma < \infty$, diverges when one requires a real tachyonic scalar field in the model.
In other words, the finite character of the effective 4d Planck mass and the gravity
localization condition on the brane imply that both $c_1$ and $c_2$ should be positive and as a consequence of this the tachyonic field
$T$ turns out to be complex in contradiction with the real nature of the starting action of the
braneworld models under consideration. We have analysed this situation for thin as well as thick branes with a
tachyonic scalar field $T$ in the bulk as in \cite{koleykar}--\cite{senguptaetal2}, with a bulk cosmological constant, 
and even for a 4d spatially flat cosmological metric induced
on the brane. We find that in all cases the tachyon field is inconsistently complex in the case when 
one requires gravity localization on this single brane super-exponential braneworld model.

Thus, it is not possible to solve the hierarchy problem in the two-brane picture and the localization of gravity in the single brane model 
with the same warp factor, contrary to the Randall-Sundrum case 
where the same warp factor is used to tackle both problems. Although, in principle, these are 
independent problems we believe it is desirable to deal with both of them within the same family of nonfactorizable 
metrics and thus the same warp factor. In this spirit, in Section \ref{canonical} we briefly discuss 
the case of a braneworld model generated by a canonical scalar field where none of these problems occur. 
Moreover, in Section \ref{tensor} we analyze the localization of 4d gravity in our braneworld by proving the existence 
of a normalizable massless zero mode on the brane. In Section \ref{stability} the stability analysis for this braneworld 
under scalar perturbations is also performed. As a result we get a stable scalar field braneworld configuration with 
no localized massive modes.

\section{The simplest tachyonic braneworld model}
\label{tbw}

We start our discussion of the problem with the simplest action we will consider and later extend it in
several ways. The action is then given as follows
\begin{equation}
S = \int d^5 x \sqrt{-g} \left(\frac{1}{2\kappa_5^2} R -V(T)\sqrt{1+g^{MN}\partial_{M} T\partial_{N} T}\right),
\label{action1}
\end{equation}
where the first term describes the 5d gravity and the second corresponds to the matter in the bulk, in this case a
$\bf{real}$ scalar tachyonic field, while $V(T)$ is its self-interacting potential \cite{sen}. 
The coefficient $\kappa_5$ is given by $\kappa^2_5=8\pi G_5$ where $G_5$ is the
5d gravitational coupling constant. The indices take the values $M,N=0,1,2,3,5$.

The 5d Einstein equations are given by
\begin{equation}
 G_{MN} = \kappa_5^2 ~T_{MN}^{\it{bulk}}.
\label{einequ}
\end{equation}
The energy--momentum tensor components read
\begin{equation}
T_{AB}^{\it{bulk}} = - g_{AB} \, V(T) \sqrt{1 +
    (\nabla T)^2} + \frac {V(T)}{\sqrt{1+ (\nabla T)^2}} \,
\partial_{A} T \, \partial_{B} T ,
\end{equation}
where $(\nabla T)^2$ is shorthand notation for $g^{MN}\partial_{M} T\partial_{N} T$. 
The covariant equation for the $T$ field is
\begin{equation}
\Box T-\frac{g^{AB}g^{MN}\nabla_A\nabla_M T\,\, \nabla_B T\,\, \nabla_N T}{1+g^{CD}\nabla_C T\,\,
\nabla_D T}= \frac{1}{V} \frac{\partial V(T) }{\partial T}. \label{fieldequ1}
\end{equation}
The background metric is given by
\begin{equation}
ds^2 = e^{-2f(\sigma)}  \eta_{\mu\nu}dx^{\mu}dx^{\nu}  + d \sigma^2 \,,
\label{metrica1}
\end{equation}
corresponding to a warped 5d line element with an induced 3--brane with a flat 4d geometry/metric.
The function $f(\sigma)$ is the warp factor and $(-,+,+,+,+)$ the signature.
Using Eq. (\ref{metrica1}) we obtain the Einstein tensor components
\begin{eqnarray}
G_{00} &=& 3\, e^{-2 f}\left( f^{''} - 2 f^{'2} \right)\,,
\label{eqeintach1}\nonumber \\
G_{i i} &=& -3\, e^{-2 f} \left(f^{''} - 2 f^{'2} \right)=-G_{00} \,,
\label{eqeintach2}
\nonumber\\
G_{\sigma\sigma} &=& 6f^{'2}\,, \label{eqeintach3}
\label{einteintensor}
\end{eqnarray}
where $i$ labels the spatial dimensions $x^i$ and a prime denotes derivative with respect to the extra dimension $\sigma$.
Since the non--diagonal components of the Einstein tensor vanish, consistency of Einstein
equations demands that non--diagonal components of the stress energy tensor should vanish
identically. This allows two possibilities: $i)$ the field $T$ depends merely on time and not
on any of the spatial coordinates --- which is the case for a scalar field in an homogeneous and
isotropic background as in cosmology; $ii)$ the field $T$ depends only on the extra dimension.  
This simply amounts to a consistent time independence of
the tachyon field even in the case when the background is time dependent, and shall be considered here. 
The tachyon field $T$ depends then only on the extra dimension $\sigma$ and Eq. (\ref{fieldequ1}) reads
\begin{equation}
T^{''}+4f^{'}T^{'}(1+T^{'2})=(1+T^{'2})\frac{\partial_{T}V(T) }{V(T)}. \label{fieldequ2}
\end{equation}
\\
The relevant Einstein equations (\ref{einequ}) can then be written as
\begin{eqnarray}
f^{''} &=& \kappa_5^2\frac{V(T)T^{'2}}{3\sqrt{1 + T^{'2}}}\,,
\label{fpp1} \\
f^{'2} &=& -\kappa_5^2\frac{V(T)}{6\sqrt{1 + T^{'2}}}. \label{fp1}
\end{eqnarray}

\noindent{\bf On the gravity localization problem.} 
In this case, we have a single brane configuration with 4d Poincar\'e symmetry 
given by Eqs. (\ref{fieldequ2}), (\ref{fpp1}) and (\ref{fp1}), and the fifth dimension is allowed to be 
infinitely large. 

It is easy to invert equations (\ref{fpp1}) and (\ref{fp1}) in terms of $T'^2$ and $V(T)$
\begin{equation}
T'^2=\frac{-f''}{2 f'^2},
\label{Tprime1}
\end{equation}
\begin{equation}
V(T)=-\frac{6 f'^2}{\kappa_5^2}\,\sqrt{1-\frac{f''}{2f'^2}}.
\label{V}
\end{equation}
For a real potential energy we obtain
\begin{equation}
f'' < 2f'^2,
\label{potreal}
\end{equation}
however a stronger condition arises from the requirement of having a real tachyonic field
\begin{equation}
f'' < 0.
\label{fppneg}
\end{equation}
For a super-exponential warp factor of the form $e^{-2f} \sim e^{-2c_1e^{c_2 |\sigma|}}$ where the warp function 
is given by
\begin{equation}
f \sim c_1 e^{c_2 |\sigma|},
\label{fexp}
\end{equation}
both the finiteness of the effective 4d Planck mass and gravity localization
conditions, which can be expressed by the following finite integral \cite{gadrr}
\begin{equation} 
\int e^{-2f(\sigma)}d\sigma < \infty, \label{Loca}
\end{equation}
require $c_1,\, c_2 >0$. Thus $f$ should be a positive definite function such that in the bulk
\begin{equation}
f'^2 \sim c_2^2 f^2 \, , \quad \quad f''\sim c_2^2 f > 0 \, .
\label{fpp}
\end{equation}
We see that $f''$ turns out to be positive, contrary to Eq. (\ref{fppneg}) which is the condition for having a real tachyonic field. 
Thus a super-exponential warp factor is not allowed in the tachyonic braneworld model described by Eq. (\ref{action1})
when one wishes to recover 4d gravity on a single brane, our world, unless we have a complex tachyonic scalar field.
\begin{table}[!htbp] \small
\begin{center}
\begin{tabular}{|@{}c@{}|@{}c@{}|@{}c@{}|@{}c@{}|}\hline
$ \quad c_1 \quad $ & $ \quad c_2 \quad $ & $ \quad f'' \quad $ & $ \quad T \quad $  \\ \hline
$ > 0 $ & $ > 0 $ & $ > 0 $ & $ \mathbb C $  \\ \hline
$ > 0 $ & $ < 0 $ & $ > 0 $ & $ \mathbb C $  \\ \hline
$ < 0 $ & $ > 0 $ & $ < 0 $ & $ \mathbb R $  \\ \hline
$ < 0 $ & $ < 0 $ & $ < 0 $ & $ \mathbb R $   \\ \hline
\end{tabular}
\end{center}
\bigskip
\caption {For the braneworld given by Eqs. (\ref{action1}) and (\ref{metrica1}) with a super-exponential warp factor (for both one- and two-brane models) we show the nature of the resulting tachyonic field $T$ in the bulk (whether real $\mathbb R$ or complex $\mathbb C$) 
depending on the chosen constant $c_1$ appearing in Eq. (\ref{fexp}). While localization of gravity on a single brane requires $c_1$ {\it and} $c_2$ positive, a real tachyonic field requires $c_1$ negative. Thus we cannot simultaneously solve both problems with the same warp factor, contrary to the Randall-Sundrum case.}
\label{table:1}
\end{table}

\noindent{\bf On the hierarchy solution within the two-brane model.} When looking at the hierarchy problem one should keep in mind the compact nature of the region of variability of $\sigma$. In this case the localization of gravity can be achieved for any sign of the constants $c_1$ and $c_2$ in the warp function $f$ since the condition (\ref{Loca}) is always fulfilled. However, in order to obtain a real $T$ field, from 
(\ref{fppneg}), (\ref{fexp}) and (\ref{fpp}) it follows that the constant $c_1$ of the warp function must be negative (see Table 1). This is true even for  the two-brane model in which the gauge hierarchy problem is solved.

We then see that we cannot simultaneously solve both the localization of gravity on a single brane and the hierarchy problem in the two-brane picture with the same warp factor, contrary to the Randall-Sundrum case. 

Although the hierarchy and the gravity localization problems are in principle independent, we believe it is desirable to deal with both of them within the same family of nonfactorizable metrics and thus the same warp factor. This has been attempted in \cite{koleykar} by using a super-exponential warp factor. While the gauge hierarchy problem has been addressed successfully with a further reduction of the hierarchy when compared to the Randall-Sundrum two-brane model result, the gravity localization problem cannot be dealt with in the single brane picture because the integral in Eq. (\ref{Loca}) becomes divergent for a real tachyonic field. 

This situation could be avoided if instead of working with a tachyonic field in the bulk we use a canonical scalar field { minimally coupled to 5d gravity as argued in Sections \ref{canonical} -- \ref{stability}  below.

\section{Extending the simplest model}

We generalise and modify our starting action in several ways, initially by introducing a 5d
cosmological constant then also by adding a thin brane, and finally by including a
spatially flat cosmological background induced on the brane. The action with a 5d
cosmological constant is given as follows
\begin{equation}
S = \int d^5 x \sqrt{-g} \left(\frac{1}{2\kappa_5^2} (R -\Lambda_5) -
V(T)\sqrt{1+g^{AB}\partial_{A} T\partial_{B} T}\right).
\label{accion2}
\end{equation}
Einstein's equations (\ref{einequ}) are now given by
\begin{eqnarray}
f^{''} &=& \kappa_5^2\frac{V(T)T^{'2}}{3\sqrt{1 + T^{'2}}}\,,
\label{einsteinequ2} \\
f^{'2} &=& -\kappa_5^2\frac{V(T)}{6\sqrt{1 + T^{'2}}}- \frac{\Lambda_5}{6}\,.
\label{restriccion2}
\end{eqnarray}
When $\sigma$ is large, the tachyon scalar field has the following behavior
\begin{equation}
T'^2=\frac{-f''}{2\left(f'^2+\frac{\Lambda_5}{6}\right)} \rightarrow \frac{-f''}{2 f'^2}\, \rightarrow 0^-.
\label{Tprime2}
\end{equation}
where the $\rightarrow$ symbol denotes large $\sigma$.
For an exponential function of the form given by Eq. (\ref{fexp}) the $f'^2$ term dominates and 
according to the last equation of (\ref{fpp}) $T$ again becomes 
complex when requiring localization of 4d gravity on a single brane.

We use now for the background metric the ansatz of a warped 5d line element with an induced 3--brane with
spatially flat cosmological background that reads
\begin{equation}
ds^2 = e^{-2f(\sigma)} \left[- d t^2 + a^2(t) \eta_{ij}dx^{i}dx^{j}  \right] + d \sigma^2 ,
\label{ansatz}
\end{equation}
while the Einstein equations (\ref{einequ}) can be rewritten in a simple way
\begin{eqnarray}
f^{''} &=& \kappa_5^2\frac{V(T)T^{'2}}{3\sqrt{1 + T^{'2}}} + e^{2 f}\, \frac{\ddot a}{a},
\label{einsteinequ} \\
f^{'2} &=& -\kappa_5^2\frac{V(T)}{6\sqrt{1 + T^{'2}}}- \frac{\Lambda_5}{6} + \frac{e^{2 f}}{2}
\left(\frac{\ddot a}{a}+\frac{\dot a^2}{a^2} \right). \label{restriccion}
\end{eqnarray}
Thus, we end up with the case of a single brane configuration in a 4d spatially
flat cosmological background given by Eqs. (\ref{fieldequ2}), (\ref{einsteinequ}) and
(\ref{restriccion}). Therefore, from these equations we conclude that we must have a de Sitter 
4d cosmological background defined by the scale factor
\begin{equation}
a(t)=e^{H\,t}, \label{scalefactor}
\end{equation}
where $H$ is the Hubble parameter and an overall constant has been absorbed into a coordinate redefinition of the 3d spatial variables \cite{gadrr}. 
This fact makes clear the role of the action of the tachyonic scalar field as a non--trivial 5d configuration which
leads to a set up in which the $dS_4$ geometry is embedded into $AdS_5$ if $\Lambda_5<0$
\cite{mannheim}. It is easy to invert these equations in terms of $T'^2$ and $V(T)$, the result
for $T'^2$ is
\begin{equation}
T'^2=\frac{-f''+H^2\,e^{2f}}{2\left(f'^2+\frac{\Lambda_5}{6}-H^2\,e^{2f}\right)} \rightarrow
\frac{H^2\,e^{2f}}{-2 H^2\,e^{2f}}=-\frac{1}{2}, \label{Tprime3}
\end{equation}
where we have taken into account relations (\ref{fpp}). Again, $T$ is a complex field, this is
easier to see for large $\sigma$ as indicated by the $\rightarrow$ symbol above. Since this 
asymptotic behaviour also holds for an arbitrary super-exponential warp factor $e^{2f}$, in principle this result
remains valid for tachyonic thick braneworld configurations.

Finally, we shall also work in conformal metric coordinates
\begin{equation}
ds^2 = e^{-2f(w)} \left[- d t^2 + a^2(t) \eta_{ij}dx^{i}dx^{j} + dw^2\right],
\label{ansatzconf}
\end{equation}
obtained from (\ref{metrica1}) with the aid of the transformation\footnote{Even when this could seem to be a simple 
coordinate play, the field equations that $T$ must obey are different for a given function $f$.} 
\begin{equation}
d\sigma=e^{-f(w)}dw.
\label{sigmaw}
\end{equation} 
Following similar steps with the aid of (\ref{fpp}), as before we get for the tachyonic field
\begin{equation}
T'^2=\frac{-f''-f'^2 + H^2+\frac{\kappa_5^2}{3}\sum_i\tau_i\delta(w-w_i)e^{-2f}}
{2\left(f'^2+\frac{\Lambda_5}{6}e^{-2f}-H^2\right)e^{2f}}
\rightarrow \frac{-f'^2}{2 f'^2 e^{2f}} =  \frac{-1}{2 e^{2f}} \rightarrow 0^-,
\label{Tprime4}
\end{equation}
where now primes stand for derivatives with respect to $w$. We have explicitly written the brane term, 
which is suppressed for large $w$. In the large $w$ (large --$f$) regime one can easily see the complex 
nature of the resultant field for any choice of the constants $c_1$ and $c_2$.

To address the hierarchy problem, where the variable $\sigma$ takes values in a finite range only, one 
should carefully construct tables similar to Table \ref{table:1} for each case under consideration and see what combinations 
of constants $c_1$ and $c_2$ give a real tachyonic field in the bulk.

\section{Canonical scalar field coupled to 5d gravity}
\label{canonical}

As we have seen the tachyon field becomes complex for large $\sigma$ in the models considered above 
when requiring the localization of 4d gravity on the brane. This problem arises because the variable 
$\sigma$ (or $w$) is of infinite range when one considers the localization of 4d gravity on the single
3--brane of the setup.
As a way out of this problem we can consider instead the action for a canonical scalar field $\phi$ minimally 
coupled to gravity plus a single 3--brane
\begin{equation}
S = \int d^5 x \sqrt{-g} \left[\frac{1}{2\kappa_5^2} (R -\Lambda_5) -
\frac{1}{2}g^{MN}\nabla_M\phi\nabla_N\phi- V(\phi)\right] + S_B,
\label{accion2fi}
\end{equation}
where $V(\phi)$ is the self--interaction potential of the bulk scalar, 
\begin{equation}
S_B = - \int d^4 x d \sigma \sqrt{-g_1} V_1(\phi) \delta(\sigma)\,,
\label{actionB}
\end{equation}
with $V_1$ the brane potential of the 3--brane 
located at the origin and endowed with an induced metric $g_1$, finally, 
we have also introduced a 5d cosmological constant. The Einstein equations with a cosmological constant in 
five dimensions are given by
\begin{equation}
 G_{MN} = -\Lambda_5 g_{MN} + \kappa_5^2 ~T_{MN}^{\it{bulk}}.
\label{einequ2fi}
\end{equation}
By using the metric Eq. (\ref{metrica1}) we obtain the components of the Einstein tensor as in Eq. 
(\ref{einteintensor}). The energy--momentum tensor components read
\begin{equation}
T_{MN}^{\it{bulk}} =  \nabla_M\phi\nabla_N\phi-
g_{MN}\left (-\frac{1}{2}g^{AB}\nabla_A\phi\nabla_B\phi- V(\phi)\right).
\end{equation}
Following the discussion after Eq. (\ref{einteintensor}) the field $\phi$ becomes a function of the fifth coordinate 
only and Einstein's equations (\ref{einequ2fi}) in the bulk are therefore written as
\begin{eqnarray}
3 f^{''} - 6 f^{'2 }&=& 
\kappa_5^2 \left( \frac{1}{2}\phi^{'2}+V  \right)+\Lambda_5\, ,
\label{ein1afi} \\
6f^{'2}&=& 
\kappa_5^2 \left( \frac{1}{2}\phi^{'2}-V  \right)-\Lambda_5,
\label{ein1cfi}
\end{eqnarray}
under the metric ansatz (\ref{metrica1}). The equation for the field $\phi$ is
\begin{equation}
\phi^{''}-4f^{'}\phi^{'}=\frac{\partial V}{\partial \phi}\,.
\label{ecampo1fi}
\end{equation}

On the other hand the junction conditions on the brane imply that the brane potential obeys
$V_1|_{\sigma=0}=6c_1c_2/\kappa_5^2\, $, implying that the brane tension is positive, while the jump of the 
first derivative of the scalar field at $\sigma=0$, denoted by $[\phi']$, 
reads $\, [\phi'] = \frac{1}{2}\frac{\partial V_1}{\partial \phi}|_{\sigma=0}$.

After some algebra we get
\begin{equation}
\phi^{'2}=\frac{3}{\kappa_5^2} f^{''},
\label{fip1}
\end{equation}
\begin{equation}
V(\phi)=\frac{3}{2\kappa_5^2}\left( f^{''}-4 f^{'2}-\frac{2}{3}\Lambda_5\right).
\label{V1fi}
\end{equation}
Thus we choose both $c_1$ and $c_2$ positive to satisfy the requirement of localization of 4d gravity given by 
Eq. (\ref{Loca}) with a {\it real} scalar field $\phi$. In this case we can consistently address both the hierarchy 
and the gravity localization problems with the same warp factor in the metric.

For the warp  function given by Eq. (\ref{fexp}) we get
\begin{equation}
\phi^{'2}=\frac{3\, c_2^2}{\kappa_5^2}f ,
\label{fip1b}
\end{equation}
\begin{equation}
V(\phi)=\frac{3\, c_2^2}{2\kappa_5^2}\left( f-4\, f^{2}-\frac{2}{3\, c_2^2}\Lambda_5\right) .
\label{V1b}
\end{equation}
The solution to Eq. (\ref{fip1b}) is
\begin{equation}
\phi = \pm 2\sqrt{\frac{3 c_1}{\kappa_5^2}} e^{\frac{c_2}{2} |\sigma|} 
= \pm 2\sqrt{\frac{3}{\kappa_5^2}} f^{1/2},
\label{fisol}
\end{equation}
from where it follows that
\begin{equation}
V(\phi)=-\frac{\Lambda_5}{\kappa_5^2}+\frac{c_2^2}{8}\phi^2-\frac{c_2^2\kappa_5^2}{24}\phi^4.
\label{Vsol}
\end{equation}

Thus, the warp factor (\ref{fexp}), the canonical scalar field (\ref{fisol}) and the self--interaction 
potential (\ref{Vsol}) define the complete solution for our scalar-tensor braneworld model.
 
Note that this potential is unbounded from below, which is common and free of pathologies when studying domain walls in $AdS_5$ supergravity \cite{dewolfe,boucher,townsend1,townsend2}. This fact was non-perturbatively established in \cite{boucher,townsend1} by determining the conditions under which the self-interaction potential $V$ of a single scalar field guarantees a stable $AdS$ vacuum, regardless of supersymmetry. It turns out that in $D$ dimensions $V$ must be of the form \cite{townsend1}
\begin{equation}
V(\phi)=2(D-2)\left[(D-2)\left(\frac{dW}{d\phi}\right)^2 - (D-1)W^2\right],
\label{adspotl}
\end{equation}
where $W(\phi)$ is the so-called superpotential, an arbitrary function of $\phi$ with at least one critical point.

Our 5D self-interaction potential (\ref{Vsol}) can be recast into the form of (\ref{adspotl}) with the aid of the superpotential 
\begin{equation}
W(\phi)=A+B\phi^2,
\label{spotl}
\end{equation}
and the following choice of the constants $$A=\pm\sqrt{\frac{\Lambda_5}{24\kappa_5^2}} \qquad {\mbox{\rm and}} \qquad B=\pm\frac{c_2\kappa_5}{24}$$ under the restriction $\Lambda_5=\frac{3c_2^2\left(1-\kappa_5^2\right)^2}{32}$. The superpotential (\ref{spotl}) clearly possesses a critical point (a maximum or a minimum depending on the sign of $B$) and then ensures the stability of the corresponding vacuum. This result is in agreement with the study of scalar perturbations that will be performed in Section \ref{stability} below.

In the next Section we shall study the stability properties of the above 
constructed braneworld model which allows for the localization of 4d gravity in contrast with previously studied 
tachyonic braneworlds with super--exponential warping in the metric \cite{koleykar}.

\section{Tensor perturbations and localization of 4d gravity}
\label{tensor}

We shall analyze the localization of 4d gravity in our braneworld field configuration following 
the work presented in \cite{Giovannini}, where a generalization of the Bardeen formalism for metric fluctuations of 
higher--dimensional backgrounds with non--compact extra dimensions was accomplished (see also \cite{KKS} and 
\cite{mg} for similar approaches); moreover, the master equations for the coupled system of metric and scalar 
perturbations were also derived in a gauge--invariant form. 

In order to analyze the dynamics of the metric fluctuations we start with the following ansatz
\begin{equation}
ds^2 = e^{-2f} \left[\left(\eta_{\mu\nu} + h_{\mu\nu}\right)dx^\mu dx^\nu + dw^2\right],
\label{metricpert}
\end{equation}
where $h_{\mu\nu}(x^\mu,w)$ are gauge--invariant metric perturbations when considering the transverse and traceless 
conditions $\partial^\mu h_{\mu\nu}=h^\mu_\mu=0$ \cite{Giovannini}. We further make use of the following separation of 
variables $h_{\mu\nu} = C_{\mu\nu} e^{3f/2} e^{ipx} \rho(w)$, where $C_{\mu\nu}$ are arbitrary constants, and the 
relevant dynamical equation adopts the form of a Schr\"odinger equation along the fifth dimension
\begin{equation}
- \rho'' + V_{QM}(w) \rho = m^2 \rho, 
\label{Schreqnrho}
\end{equation}
where $m$ is the mass that a 4d observer sees \cite{csakietal}, whereas the effective quantum mechanical potential 
reads
\begin{equation}
V_{QM}(w) = \frac{s''}{s} \equiv {\cal J}^2 - {\cal J}', \qquad \mbox{\rm with} \qquad s = e^{-3f/2}.
\label{potVQMrho}
\end{equation}
In this equation we have introduced the quantity ${\cal J} = -\frac{s'}{s}$, called superpotential within the 
framework of supersymmetric quantum mechanics. Moreover, this superpotential allows us to express the Schr\"odinger 
equation for $\rho$ (\ref{Schreqnrho}) as follows
\begin{equation}
{\cal Q}^{\dagger} {\cal Q}\ \rho = m^2 \rho,
\label{susyqmrho}
\end{equation}
where the operators ${\cal Q}^{\dagger}$ and ${\cal Q}$ are defined according to
\begin{equation}
{\cal Q}^{\dagger} = \biggl( - \frac{d}{dw} + {\cal J} \biggr), \qquad\qquad
{\cal Q} = \biggl( \frac{d}{dw} + {\cal J} \biggr).
\label{operators}
\end{equation}
The fact that the Schr\"odinger equation (\ref{Schreqnrho}) can be expressed in the form (\ref{susyqmrho}) guarantees
that the spectrum of metric fluctuations is positive definite and there are no tachyonic modes with $m^2 <0$, 
guaranting the stability of the system under the tensorial sector of perturbations. 

For the massless zero mode the Schr\"odinger equation (\ref{Schreqnrho}) yields $\rho=s=e^{-3f/2}$ with the following
normalization condition 
\begin{equation}
\int e^{-3f(w)}dw < \infty, 
\label{Locaw}
\end{equation}
which transforms into (\ref{Loca}) when performing the coordinate transformation (\ref{sigmaw}). Since this integral
converges for positive $c_1$ and $c_2$, this guarantees the existence of a normalizable massless zero mode which
is interpreted as a 4d graviton localized on the brane.

\section{Stability of the brane under scalar perturbations}
\label{stability}

In this Section we shall perform the stability analysis of our braneworld configuration under linear perturbations 
of the scalar sector following again the line of work of \cite{Giovannini}, where the master equations
for the coupled system of scalar perturbations were derived. 

Thus we shall consider spin--$0$ linear perturbations of the scalar-gravity coupled system that generates our
braneworld and study their dynamics as well as the structure of the corresponding mass spectra. Although in the 
previous Section we used a gauge--invariant treatment, in the scalar stability analysis it is more convenient to 
use a specific gauge.

Let us start by considering the perturbed metric for the scalar sector of fluctuations written in conformal 
coordinates in the so--called longitudinal gauge 
\begin{equation}
ds^2 = e^{2f(z)}\left[(1+2\psi)\eta_{\mu\nu}dx^{\mu}dx^{\nu}+(1+2\xi)dw^2\right],
\end{equation}
together with the fluctuations of the scalar field $\varphi=\phi+\chi$, where $\psi$, $\xi$ and $\chi$ are small 
perturbations.

In \cite {Giovannini} it was shown that the corresponding system of coupled differential equations can be reduced to 
a couple of master equations which can be suitable expressed in a Schr\"odinger--like form. Moreover, it was 
determined that there is only one independent degree of freedom, i.e. just one scalar physical mode, after taking 
into account the following relations $\xi=2\psi$ and $\chi=-\frac{6}{\phi'}\left(\psi'-2f'\psi\right)$.  

Therefore the equation for the rescaled scalar perturbation $\psi$ can be recast into the Schr\"odinger form 
\begin{equation}
\Psi'' - g \left(g^{-1}\right)'' \Psi = m_{\Psi}^2\Psi ,
\label{Psieqn}
\end{equation}
after a convenient separation of 4d and 5d variables which define the 4d mass $m_{\Psi}$ and by setting 
$\Psi=\frac{e^{-3f/2}}{\phi'}\psi$ (a similar equation holds for the rescaled scalar fluctuation $\xi$). In this 
equation it was introduced a very useful function 
\begin{equation}
g = \frac{e^{-3f/2}\phi'}{f'} ,
\label{g}
\end{equation}
that parameterizes the analogue quantum mechanical potential
\begin{equation}
U_{\Psi} = g \left(g^{-1}\right)''. 
\label{VQMPsi}
\end{equation}
On the other hand, the corresponding equation for the scalar perturbation $\chi$ also transforms into a
Schr\"odinger--like equation
\begin{equation}
X'' - g^{-1}g'' X = m_X^2 X,
\label{Xeqn}
\end{equation}
after the definition of the new fluctuation $X=e^{-3f/2}\chi-g\psi$ and the introduction of the 4d mass
of this scalar perturbation $m_X$, where now the analogue quantum mechanical potential reads
\begin{equation}
U_{X} = g^{-1}g''. 
\label{VQMX}
\end{equation}
These potentials can respectively be written in the following form 
\begin{equation}
U_{\Psi} = {\cal J}_{\Psi}^2 - {\cal J}_{\Psi}', \qquad\qquad U_X = {\cal J}_X^2 + {\cal J}_X', 
\label{pots}
\end{equation}
with the aid of the superpotentials ${\cal J}_{i} = \frac{g'}{g}$, with $i=\Psi,X$. As in the case of metric 
fluctuations, these quantities allow us to write the Schr\"odinger equations for $\Psi$ and $X$ as
\begin{eqnarray}
&& {\cal Q}^{\dagger} {\cal Q} \Psi = m^2_{\Psi} \Psi,
\nonumber\\
&&   {\cal Q} {\cal Q}^{\dagger} X = m^2_X X,
\end{eqnarray}
where the operators ${\cal Q}^{\dagger}$ and ${\cal Q}$ are defined according to (\ref{operators}).

Again, since the Schr\"odinger equation for both $\Psi$ and $X$ can be written in this form, this means that 
there are no tachyonic modes with $m_i^2 <0$ in the respective spectra of scalar fluctuations and thus the 
scalar-tensor braneworld configuration that defines our braneworld model is stable under linear scalar perturbations.

\subsection{Delocalization of the scalar modes}

In order to study the (de)localization properties of the scalar perturbation sector it is enough to understand the 
behaviour of the analogue quantum mechanical potentials (\ref{VQMPsi}) and (\ref{VQMX}) of the corresponding 
Schr\"odinger equations (\ref{Psieqn}) and (\ref{Xeqn}). These potentials are written in terms of the 
conformal coordinate $w$ while our solution is parameterized in the language of the coordinate $\sigma$.
However, the dependence of these potentials can be plotted parametrically once we have obtained 
$w=\left[\mbox{\rm Ei}\left(-c_1e^{c_2|\sigma|}\right)-\mbox{\rm Ei}\left(-c_1\right)\right]\mbox{\rm sgn}(\sigma)/c_2
=\left[\mbox{\rm Ei}\left(-f\right)-\mbox{\rm Ei}\left(-c_1\right)\right]\mbox{\rm sgn}(\sigma)/c_2$, where 
$\mbox{\rm Ei}(x)$ stands for the exponential integral special function, with the aid of the coordinate transformation 
(\ref{sigmaw}). 

It turns out that for our superexponential warp factor (\ref{fexp}) and the corresponding background scalar field 
(\ref{fisol}), both $U_i(w)$ are positive potential barriers distributed along the fifth dimension, a fact that 
implies that no massless nor massive scalar modes are localized on the brane within our model.

\section{Discussion and conclusions}

The finiteness of the 4d effective gravitational coupling constant and the gravity 
localization conditions, which are encoded in the requirement
that $\int e^{-2f(\sigma)}d\sigma < \infty$, imply that both $c_1$ and $c_2$ should be positive 
for a tachyonic braneworld model with super-exponential warp factors of the form $e^{-2f} \sim e^{-2c_1 e^{c_2 |\sigma|}}$.
As a consequence of this the examples considered within this context 
show that the tachyonic field $T$ turns out to be complex 
in contradiction with the real
character of the starting tachyonic action. The inverse statement is also valid: if the tachyonic scalar 
field is required to be real then the 4d gravity is not localized on the brane and the effective 4d Planck 
mass turns out to be infinitly large. 
We have analysed this situation for thin as well as thick braneworlds generated
by gravity and a tachyonic field $T$, with an additional bulk cosmological constant, and even for
a spatially flat 4d cosmological metric induced on the brane and shown that in all
cases the tachyon field is inconsistently complex if 4d gravity is required to be 
localized on a single 3--brane.

On the other hand there has recently been some interest in using {\it super-exponential} warp factors to address the hierarchy and localization problems in tachyonic braneworld type models. As originally shown by Randall and Sundrum and by Gogberashvili the hierarchy problem of masses in the standard model of particles can be considerably ameliorated by introducing a
factor of the form $e^{-2f(\sigma)}$ corresponding to a warped 5d line element with an induced
3--brane in such a way that higher scales are exponentially suppressed. Thus scales of the order
of the Planck mass end up with values a few tens the electroweak scale. This is achieved by using
warp functions of the type $f \sim k|\sigma|$. A further reduction of the hierarchy was acomplished in 
\cite{koleykar} with the aid of a super-exponential warp factor with $f \sim c_1 e^{c_2 |\sigma|}$ which leads to a model with physical parameters of the same order. In general care should be taken when chosing the constants $c_1$ and $c_2$ because 
the wrong choice can result in an imaginay tachyonic field $T$ in some braneworld models of this type. 
We also find that it is not possible to solve the hierarchy problem {\it and} the localization of gravity with the same warp factor within these models. This is contrary to the Randall-Sundrum model case where the same warp factor is used in both problems. 

The gauge hierarchy and the 4d gravity localization problems are independent 
but since both of them occur in the same universe we consider desirable to 
solve both of them with the same nonfactorizable metric and therefore the same warp factor.
With this motivation we proposed considering a bulk canonical scalar field instead of a tachyonic one in order to
generate a braneworld model and show that the above mentioned problems can be solved with the same 
super--exponential warp factor. Finally, this latter scalar braneworld configuration is shown to be stable
under scalar linear fluctuations. Moreover, the scalar modes, both massless and massive, were shown not to be localized on the brane.

It would be interesting to further study the dynamics of the massive KK tensorial fluctuations in the single brane picture and see how do they correct the Newton's law, for instance. However, this task is not trivial at all from the mathematical point of view and it seems that numerical tools should be implemented to afford it. Notwithstanding, numerical studies render qualitatively results and do not allow one to obtain precise predictions about these corrections within the braneworld paradigm (see \cite{bhknq} and references therein for an example on this issue).

\section*{\bf Acknowledgements}

We gratefully acknowledge support from \textquotedblleft
Programa de Apoyo a Proyectos de Investigaci\'on
e Innovaci\'on Tecnol\'ogica\textquotedblright\, (PAPIIT) UNAM, IN103413-3, {\it Teor\'ias de Kaluza-Klein, 
inflaci\'on y perturbaciones gravitacionales.} AHA is grateful to the staff of ICF, UNAM and MCTP, UNACH for hospitality. 
MCG and DMM acknowledge a scholarship and a posdoctoral fellowship from DGAPA-UNAM, respectively. 
GG and AHA thank SNI for support.


\begin{thebibliography}{999}

\bibitem{gogrs} M. Gogberashvili, (1999).  Four dimensionality in noncompact Kaluza-Klein
model, {\it Mod. Phys. Lett.} {\bf A14} 2025, hep-ph/9904383;  (2002). Hierarchy problem in the shell
universe model, {\it Int. J. Mod. Phys.} {\bf D11} 1635,  hep-ph/9812296;
L. Randall and R. Sundrum, (1999). A large mass hierarchy from a small extra dimension,  {\it Phys.
Rev. Lett.} {\bf 83} 3370, hep-ph/9905221; (1999). An alternative to compactification, {\it Phys.
Rev. Lett.} {\bf 83} 4690, hep-th/9906064.

\bibitem{thbrs} V. Dzhunushaliev, V. Folomeev and M. Minamitsuji, (2010). Thick brane solutions,
{\it Rept. Prog. Phys.} {\bf 73} 066901, arXiv:0904.1775 [gr-qc].

\bibitem{sengupta} S. SenGupta, (2008). Aspects of warped braneworld models, arXiv:0812.1092 [hep-th].



\bibitem{mannheim} P.D. Mannheim, {\it Brane--Localized Gravity}, World Scientific (2005).

\bibitem{maartens} R. Maartens and K. Koyama, (2010). Brane--World Gravity,  {\it Living Rev. Rel.} {\bf 13} 5,
arXiv:1004.3962 [hep-th].

\bibitem{dewolfe} O. De Wolfe, D.Z. Freedman, S.S. Gubser and A. Karch, (2000). Modeling
the fifth dimension with scalars and gravity, {\it Phys. Rev.} {\bf D62} 046008, hep-th/9909134.

\bibitem{gremm} M. Gremm, (2000). Four--dimensional gravity on a thick domain wall,
{\it Phys. Lett.} {\bf B478} 434, hep-th/9912060;  (2000). Thick domain walls and singular spaces,
{\it Phys. Rev.} {\bf D62} 044017, hep-th/0002040.

\bibitem{csakietal} C. Csaki, J. Erlich, T. Hollowood and Y. Shirman, (2000). Universal
Aspects of gravity localized on thick branes, {\it Nucl. Phys.} {\bf B581} 309, hep-th/0001033.

\bibitem{Giovannini} M. Giovannini, (2001). Gauge--invariant fluctuations of scalar branes, {\it Phys. Rev.} {\bf D64} 064023, hep-th/0106041; (2002). Localization of metric fluctuations on scalar branes, {\it Phys. Rev.} {\bf D65} 064008, hep-th/0106131. 

\bibitem{nonmincoupling} K. Farakos and P. Pasipoularides, (2007). Gauss--Bonnet
gravity, brane world models, and non--minimal coupling, {\it Phys. Rev.} {\bf D75} 024018,
hep-th/0610010; K. Farakos, G. Koutsoumbas and P. Pasipoularides, (2007). Graviton
localization and Newton's law for brane models with a non-minimally coupled bulk scalar
field, {\it Phys. Rev.} {\bf D76} 064025, arXiv:0705.2364 [hep-th]; A. Herrera--Aguilar, D. Malag\'on--Morej\'on, R.R.
Mora--Luna and I. Quiros, (2012). Thick braneworlds generated by a non-minimally coupled scalar field
and a Gauss-Bonnet term: conditions for localization of gravity, {\it Class. Quantum Grav.} {\bf 29} 035012, arXiv:1105.5479 [hep-th]; H. Guo, Y.--X. Liu, Z.--H. Zhao and F.--W. Chen, (2012). Thick branes with a non--minimally coupled bulk--scalar field,  {\it Phys. Rev.} {\bf D85} 124033, arXiv:1106.5216 [hep-th].

\bibitem{KKS}
S. Kobayashi, K. Koyama and J. Soda, (2002). Thick brane worlds and their stability, {\it Phys.
Rev.} {\bf D65} 064014, hep-th/0107025.

\bibitem{mg}
S. Mert Aybat and D.P. George, (2010). Stability of scalar fields in warped extra dimensions, {\it JHEP} {\bf 1009} 
010, arXiv:1006.2827 [hep-th].

\bibitem{fRwaveBW} 
V.I. Afonso, D. Bazeia, R. Menezes and A.Yu. Petrov, (2007). f(R)-Brane, {\it Phys. Lett.} {\bf B658} 71, arXiv:0710.3790 [hep-th]; Y. Zhong, Y.--X. Liu and K. Yang, (2011). Tensor perturbations of f(R)--branes, {\it Phys. Lett.} {\bf B699} 398,
arXiv:1010.3478 [hep-th]; M. Gogberashvili and D. Singleton, (2010). Anti-de-Sitter Island-Universes from 5D Standing Waves, {\it Mod. Phys. Lett.} {\bf A25} 2131, arXiv:0904.2828 [hep-th]; M.
Gogberashvili, A. Herrera--Aguilar and D. Malag\'on--Morej\'on, (2012). An Anisotropic Standing Wave
Braneworld and Associated Sturm-Liouville Problem, {\it Class. Quantum Grav.} {\bf 29} 025007,
arXiv:1012.4534 [hep-th].

\bibitem{ariasybarbosaetal} O. Arias, R. Cardenas and Israel Quiros, (2002). Thick brane worlds arising from
pure geometry, {\it Nucl. Phys.} {\bf B643} 187, hep-th/0202130; N. Barbosa--Cendejas and A. Herrera--Aguilar, (2005). 4D gravity localized in non $Z_2$--symmetric thick branes, {\it JHEP} {\bf 10} 101; hep-th/0511050; (2006). Localization of 4D gravity on pure
geometrical thick branes, {\it Phys. Rev.} {\bf D73} 084022; (2008). {\it Erratum-ibid.} {\bf D77} 049901, hep-th/0603184; N. Barbosa--Cendejas, A. Herrera--Aguilar, M.A. Reyes and C. Schubert, (2008). Mass gap for gravity localized on Weyl thick branes, {\it Phys. Rev.} {\bf D77} 126013, arXiv:0709.3552.

\bibitem{SR} J.M. Hoff da Silva and R. da Rocha, (2009). Braneworld remarks in Riemann--Cartan manifolds,
{\it Class. Quantum Grav.} {\bf 26} 055007; (2009). {\it Erratum--ibid.} {\bf 26} 179801,
arXiv:0804.4261 [gr-qc]; (2010). Torsion Effects in Braneworld Scenarios, {\it Phys. Rev.}
{\bf D81} 024021, arXiv:0912.5186 [hep-th]; (2010). Gravitational constraints of
dS branes in AdS Einstein--Brans--Dicke bulk, {\it Class. Quantum Grav.} {\bf 27} 225008,
arXiv:1006.5176 [gr-qc]; (2012). Effective Monopoles within Thick Branes, {\it Europhys. Lett.} {\bf 100}  11001, arXiv:1209.0989 [hep-th].

\bibitem{Bazeiaetal} D. Bazeia, F.A. Brito and J.R. Nascimento, (2003). Supergravity brane worlds and tachyon potentials, {\it Phys.
Rev.} {\bf 68} 085007, hep-th/0306284.

\bibitem{koleykar} R. Koley and S. Kar, (2005). A Novel braneworld model with a bulk scalar field,
{\it Phys. Lett.} {\bf B623} 244, (2005). {\it Erratum--ibid.} B{\bf 631} 199, hep-th/0507277.

\bibitem{senguptaetal1} D. Maity, S. SenGupta, and S. Sur, (2006). Stability analysis of the Randall-Sundrum
braneworld in presence of bulk scalar, {\it Phys. Lett.} {\bf B643} 348, hep-th/0604195.

\bibitem{palkar} S. Pal and S. Kar, (2009). de Sitter branes with a bulk scalar, {\it Gen. Rel. Grav.} {\bf 41} 1165,
hep-th/0701266.

\bibitem{senguptaetal2} A. Das, S. Kar and S. SenGupta, (2009). Stable two--brane models with bulk tachyon matter,
{\it Int. J. Mod. Phys.} {\bf A24} 4457, arXiv:0804.1757 [hep-th].

\bibitem{koley} R. Koley, (2008). Localization of fields on brane, in {\it Proceedings of the Workshop on ``Physics
of warped extra dimensions"} 19, arXiv:0812.1423 [hep-th]; R. Koley, J. Mitra and S.
SenGupta, (2009). Fermion localization in generalised Randall Sundrum model, {\it Phys. Rev.} {\bf D79} 041902, arXiv:0806.0455 [hep-th].

\bibitem{zld} X.H. Zhang, Y.X. Liu, and Y.S. Duan, (2008). Localization of Fermionic Fields on Braneworlds with
Bulk Tachyon Matter, {\it Mod. Phys. Lett.} {\bf A23} 2093, arXiv:0709.1888 [hep-th].

\bibitem{gadrr} G. Germ\'an, A. Herrera-Aguilar, D. Malag\'on-Morej\'on, R.R. Mora-Luna and 
R. da Rocha, (2013). A de Sitter tachyon thick braneworld and gravity localization, {\it JCAP} {\bf 1302} 035, 
arXiv:1210.0721 [hep-th].

\bibitem{sen} 
A. Sen, (1999). Supersymmetric world volume action for nonBPS D--branes, {\it JHEP} {\bf 9910} 008, hep-th/9909062; 
E.A. Bergshoeff, M. de Roo, T.C. de Wit, E. Eyras, Sudhakar Panda, (2000). T duality and actions for non-BPS D-branes, {\it JHEP} {\bf 0005} 009, arXiv:hep-th/0003221; A. Sen, (2002). Rolling tachyon, {\it JHEP} {\bf 0204} 048, hep-th/0203211; (2002). Tachyon matter, {\it JHEP} {\bf 0207} 065, hep-th/0203265; (2002). Field theory of tachyon matter, {\it Mod. Phys. Letts.} {\bf A17} 1797, hep-th/0204143; (2003). Time and tachyon, {\it Int. J. Mod. Phys.} {\bf A18} 4869, hep-th/0209122.

\bibitem{boucher}  
W. Boucher, (1984). Positive energy without supersupersymmetry, {\it Nucl. Phys.} {\bf B242} 282. 

\bibitem{townsend1} 
P.K. Townsend, (1984). Positive energy and the scalar potential in higher dimensional (super)gravity theories, 
{\it Phys. Lett.} {\bf B148} 55. 

\bibitem{townsend2} 
K. Skenderis and P.K. Townsend, (1999). Gravitational stability and renormalization--group flow, 
{\it Phys. Lett.} {\bf B468} 46, hep--th/9909070.

\bibitem{bhknq} 
N. Barbosa-Cendejas, A. Herrera--Aguilar, K. Kanakoglou, U. Nucamendi and I. Quiros, (2013). Mass gap, mass hierarchy and corrections to Newton's law on thick branes with Poincar\'e symmetry, to appear in {\it Gen. Rel. Grav.} 

\end{thebibliography}
\end{document}